\documentstyle[twoside,fleqn,espcrc2,epsf]{article}

\newcommand{\AmS}{{\protect\the\textfont2
  A\kern-.1667em\lower.5ex\hbox{M}\kern-.125emS}}

\title{Confinement, Chiral Symmetry Breaking and Continuum Limits \\
in Quantum Link Models}

\author{Shailesh Chandrasekharan
	\address{Department of Physics, Duke University,
        Durham, NC 27708}%
        \thanks{This work was supported in part by US department of
		energy grant DE-FG02-96ER40945.}
     }  
\begin{document}

\begin{abstract}
Using the example of compact $U(1)$ lattice gauge theory we argue that 
quantum link models can be used to reproduce the physics of conventional
Hamiltonian lattice gauge theories. In addition to the usual gauge 
coupling $g$, these models have a new parameter $j$ which naturally 
cuts-off large electric flux quanta on each link while preserving exact 
$U(1)$ gauge invariance. The $j\rightarrow\infty$ limit recovers the 
conventional Hamiltonian. At strong couplings, the theory shows confinement 
and chiral symmetry breaking for all non-trivial values of $j$. The phase 
diagram of the 3+1 dimensional theory suggests that a coulomb phase 
is present at large but finite $j$. Setting $g=0$, a new approach to the 
physics of compact $U(1)$ gauge theory on the lattice emerges. In this
case the parameter $j$ takes over the role of the gauge coupling, and 
$j\rightarrow \infty$ describes free photons.
\end{abstract}
%\hfill DUKE-TH-98-174
\maketitle

\section{INTRODUCTION}

Hamiltonian lattice gauge theories were proposed a long time ago\cite{Kogut}. 
Unfortunately, a class of extensions\cite{Orland,Sch}, which we refer to as 
quantum link models, were left unexplored. These models suggest an
interesting alternative for tackling gauge theories. For example they 
give new insight into cluster algorithms for gauge theories\cite{Bernard}. 
For a review of the subject look at \cite{Uwe}. 

An interesting question which we wish to address here is how the new 
formulations can reproduce the physics of the conventional formulations. We 
consider this question in the context of compact $U(1)$ gauge theory. 
In particular, we show that the new formulations show confinement and chiral 
symmetry breaking at strong couplings. At weak couplings, we argue that 
these models can have a coulomb phase and thus can also describe massless 
photons. The presence of these two different phases shows that quantum link 
models can also exhibit rich and complex dynamics.

The conventional Kogut-Susskind Hamiltonian for $U(1)$ lattice gauge 
theory is given by
\begin{equation}
\label{eq:KSH}
H = \frac{g^2}{2}\sum_{x,\mu} E^2_{x,\mu} -
\frac{1}{2g^2}\sum_{\cal P} (U_{\cal P} + U^\dagger_{\cal P})
\end{equation}
where $E_{x,\mu}$ and $U_{x,\mu}$ are operators on each link $(x,\mu)$.
$U_{\cal P}$ is the plaquette operator given by 
$U_{\cal P} = U_{x,\mu}U_{x+\mu,\nu}U^\dagger_{x+\nu,\mu}U^\dagger_{x,\nu}$
The electric field operator $E$ and the gauge field operator $U$ on each
link obey the commutation relations,
$[E,U] = U$, $[E,U^\dagger] = -U^\dagger$, and $[U,U^\dagger] = 0$.
These commutation relations cannot be realized by operators in a finite
dimensional space. For example, the eigenvalues $m$ of $E$ are integers, 
$m= -\infty...,-1,0,1,...\infty$. Further, $U$ also can be diagonalized, 
although not simultaneously with $E$. Its eigenvalues, $\exp(iA)$ with 
$A = (-\pi,\pi]$, are the usual $U(1)$ phases.

However, it is not necessary to insist that $[U,U^\dagger]=0$ to preserve
the gauge invariance of the Hamiltonian, since it is the commutation 
relations between $E$ and $U$ or $U^\dagger$ that make $H$ gauge invariant. 
In particular by demanding that $[U,U^\dagger] = 2E/[j(j+1)]$, for a 
positive integer $j$, we can extend the Kogut-Susskind Hamiltonian to a 
class of gauge invariant Hamiltonians $H^{(j)}$ dependent on a new parameter 
$j$. In the $j\rightarrow \infty$ limit the conventional Hamiltonian $H$ is 
regained. For finite $j$, $E$ and $U$ can be represented by spin-$j$ operators
and we will call them $E^{(j)}$ and $U^{(j)}$ instead. $E^{(j)}$ has 
eigenvalues between $-j$ and $j$ and $U^{(j)}$ is no longer diagonalizable 
for finite $j$.  We will argue below that, even with these constraints
interesting qualitative features of the theory are reproduced at finite $j$.

\section{STRONG COUPLINGS}

The new class of Hamiltonians $H^{(j)}$ are formulated on a Hilbert space 
where the magnitude of the electric flux is bounded by $j$ units on each 
link. At strong couplings, since large electric flux states are 
naturally suppressed, $H^{(j)}$ and $H$ will describe qualitatively similar
physics. The dynamics of conventional lattice gauge theories at strong 
couplings produce confinement and chiral symmetry breaking. We illustrate 
how these phenomena emerge in quantum link models using the (1+1) dimensional 
lattice Schwinger model constructed with staggered fermions. Similar results 
in higher dimensions can also be derived with some work.

Lattice Schwinger model with staggered fermions in the Hamiltonian 
formulation at strong couplings was first studied by Banks et.al., and
a more accurate analysis was presented recently by Berruto et.al.,
\cite{Berruto}. We extend the latter results to the quantum link model
whose Hamiltonian is given by
\begin{eqnarray}
&& H^{(j)}_{\rm Schw} =
\sum_{x=0}^{N-1} \left[ \frac{g^2a}{2} 
(E^{(j)}_x - (-1)^x\frac{1}{4})^2 \right. 
\nonumber \\
&& \left. \hbox{\hskip0.6in} \;-\; 
(-1)^x m (\psi^\dagger_x\psi_x - \frac{1}{2}) \right]
\nonumber \\
&& \;-\; 
\frac{it}{2a} \sum_{x=0}^{N-1} \left( \psi^\dagger_{x+1}U^{(j)}_x\psi_x - 
\psi^\dagger_x (U^{(j)})^\dagger_x\psi_{x+1}\right)
\nonumber
\end{eqnarray}
where N is even. Ignoring the quantum link nature of $E^{(j)}$ and 
$U^{(j)}$, the model has been studied by Berruto et.al. We refer the
readers to their article for a more complete discussion and justifications
for the above form of the Hamiltonian. The gauge generator
\begin{equation}
G_x = E_x - E_{x-1} + \psi^\dagger_x\psi_x - \frac{1}{2}(1+(-1)^x)
\end{equation}
can be used to define gauge invariant states in the Hilbert space. 

The strong coupling ground state energy of the Hamiltonian is given by 
$E_0 =  Ng^2a/32 - m N/2$. The energy $E$ of a state with two heavy
charges separated by a distance of $R$,  is given by $(E-E_0)/(g^2a) =  
2m/(g^2a) + R + 1/2$, which shows the linear confining potential. It is
possible to obtain the pseudo-scalar($\pi$) and scalar($\sigma$) masses 
in the strong coupling expansion. If the expansion parameter is taken to be
$\epsilon = t/2g^2a^2$, and $\mu=m/g^2a$, up to fourth order we get
\begin{eqnarray}
\frac{E_{\pi}}{g^2a} 
&=& \frac{E_0}{g^2a} + \frac{1}{4} + 2\mu  + \epsilon^2 \frac{8}{1+8\mu} 
\nonumber \\
&-& \epsilon^4 
\frac{512 + p(j)(64 + 512\mu)}{(1+8\mu)^3} 
\nonumber \\
\frac{E_{\sigma}}{g^2a} 
&=& \frac{E_0}{g^2a} + \frac{1}{4} + 2\mu  + \epsilon^2 \frac{24}{1+8\mu} 
\nonumber \\
&-& \epsilon^4 \frac{1536 + p(j)(64 + 512\mu)}{(1+8\mu)^3} 
\nonumber \\
\frac{E_0}{g^2a} &=& \frac{N}{32} - \frac{\mu N}{2} - 
\epsilon^2 \frac{4N}{1+8\mu}
 + \epsilon^4 \frac{192N}{(1+8\mu)^3} 
\end{eqnarray}
The value of $j$ begins to enter only at the fourth order in the expansion, 
through $p(j) = (j-1)(j+2)/[j(j+1)]$, which clearly shows the minor role 
of $j$. Chiral symmetry is also broken, and can be seen by a non-zero value 
of the condensate. Up to fourth order
\begin{equation}
\langle \bar\psi\psi \rangle =  -\frac{1}{2a}
\left[ 1 - \epsilon^2\frac{32}{(1+8\mu)^2}
+ \epsilon^4\frac{1536}{(1+8\mu)^4} \right]
\end{equation}
which shows that now $j$ enters the result only at higher orders.
In the $j\rightarrow \infty$ limit, the above results match with 
\cite{Berruto}. On general grounds it can be argued that the minor role 
of $j$ at strong couplings is a universal feature and extends to higher 
dimensions.

\section{COULOMB PHASE}

\begin{figure}[htb]
\vspace{9pt}
%\framebox[55mm]{\rule[-21mm]{0mm}{43mm}}
\epsfxsize=70mm
\epsffile{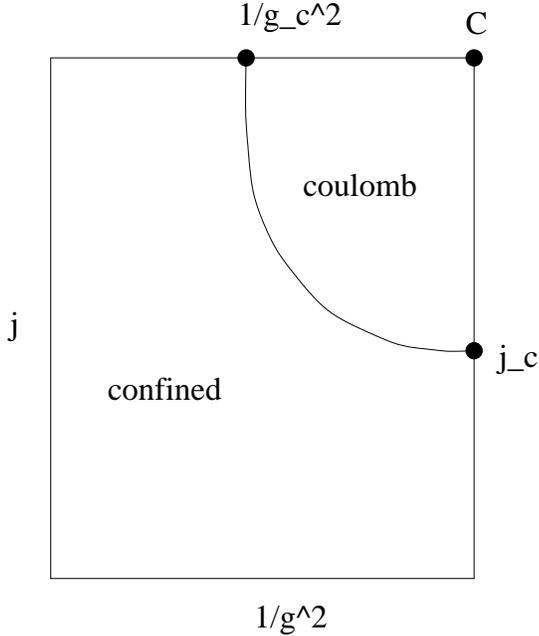}
\caption{The conjectured phase diagram of 3+1 dimensional $U(1)$ quantum 
link model. The point C refers to the theory with free photons.}
\label{fig:phase}
\end{figure}

  In 3+1 dimensions the conventional compact $U(1)$ gauge theory without 
matter fields has interesting phase structure. The Hamiltonian is given 
by eq.(\ref{eq:KSH}). The theory has a confined phase at strong couplings 
and a coulomb phase at weak couplings. By extending the Hamiltonian to 
$H^{(j)}$, we can study the phase diagram in the $j$ vs. $1/g^2$ plane. 
Based on results from the previous section we can expect that the strong 
coupling phase will extend to small $j$. On the other hand, disallowing 
large electric flux states can naturally make the theory strongly coupled 
even at small $g$. Thus the weak coupling phase may not be accessible at
small $j$. A simple plausible phase diagram is given in figure. 
{\ref{fig:phase}}

The point C represents the fixed point describing free photons and
usually approached along the $j= \infty$ axis in the conventional
formulation. However, the phase diagram, if correct suggests that we can 
also approach it along the $g = 0$ axis. Setting $g=0$ in $H^{(j)}$ we 
obtain the quantum link Hamiltonian in $d$ spatial dimensions\cite{Sch}, 
given by
\begin{equation}
\label{eq:QLH}
H^{(j)}(g=0) \;=\;\; \sum_{\cal P}\;\;\; [\;U^{(j)}_{\cal P} \;+\; 
(U^{(j)}_{\cal P})^\dagger\;]
\end{equation}
where $j$ is the only free parameter. We expect that in the 
$j\rightarrow \infty$ limit this Hamiltonian describes free photons. 
To see this let us consider the large $j$ expansion. We first write
\begin{equation}
U = \left(1 + \frac{E}{j(j+1)} - 
\frac{ E^2}{j(j+1)}\right)^{\frac{1}{2}} {\rm e}^{iA}
\end{equation}
where, $[A,E] = i$, which reproduces the necessary commutation relations.
We then substitute $U$ in eq.(\ref{eq:QLH}) to obtain a low energy, large 
$j$ effective (tree-level)Hamiltonian,
\begin{eqnarray}
H_{\rm eff} &=& \frac{2}{\sqrt{j(j+1)}}\left(
\frac{c^2}{2\sqrt{j(j+1)}}\sum_{x,\mu} E^2_{x,\mu} \right.
\nonumber \\
&& \;\;\;\;\;\;\;\;\;\;\;\;\;\;\;\;\;
+ \left. \frac{\sqrt{j(j+1)}}{2} \sum_{x,\mu}B^2_{x,\mu}\right)
\end{eqnarray}
where $B$ is the magnetic field with the usual definition and 
$c^2 = 2(d-1)$ is the speed of light. $H_{\rm eff}$ shows that $j$ takes 
over the role of the coupling through the relation 
$g_{\rm eff}^2 \sim 1/\sqrt{j(j+1)}$. As expected smaller $j$ makes the 
theory effectively strongly coupled. The results of the large $j$ expansion 
gives further credibility to the conjecture of the phase diagram shown in 
figure \ref{fig:phase}. Since $g_c \sim 1$ in the conventional formulation, 
based on the tree level relation between $g_{\rm eff}$ and $j$, we may 
expect that $j_c$ is of order one. Perhaps $j \sim 1$ or $2$  will already 
describe a coulomb phase! The presence of such a phase shows the 
potential of quantum link models to describe long range physics and continuum 
limits. 

\section{Acknowledgment}
I would like to thank Uwe Wiese for his kind hospitality during my visit to 
MIT and lots of discussions on the subject. I would also like to thank 
B. Beard, R. Brower, A. Tsapalis for many useful discussions.

\end{document}